\documentclass[sigconf,twocolumn,table]{acmart}

\date{}

\AtBeginDocument{%
  }

\setcopyright{acmlicensed}
\setcopyright{cc} 
\setcctype{by} 
\copyrightyear{2025} 
\acmYear{2025} 
\acmDOI{10.1145/3772052.3772226}

\acmConference[SoCC '25]{ACM Symposium on Cloud Computing}{November 19--21, 2025}{Online, USA} 
\acmBooktitle{ACM Symposium on Cloud Computing (SoCC '25), November 19--21, 2025, Online, USA}
\acmISBN{979-8-4007-2276-9/2025/11}

\usepackage[]{hyperref}

\usepackage{tikz}
\usepackage{amsmath}
\usepackage{comment}
\usepackage{grumble}
\usepackage{xspace}

\usepackage{threeparttable} 

\usepackage{pifont} 
\newcommand{\cmark}{\ding{51}}%
\newcommand{\xmark}{\ding{55}}%

\usepackage{filecontents}

\usepackage{libertine}

\usepackage{enumitem}
\usepackage{tabu}

\usepackage{array}
\usepackage{multirow}
\usepackage{arydshln}
\newcolumntype{P}[1]{>{\centering\arraybackslash}p{#1}}

\usepackage{soul}

\newcommand{\myparagraph}[1]{\smallskip \noindent{\bf {#1}.}}
\newcommand{\myidea}{\noindent{\bf Key idea.}}

\newcommand{\out}[1] {}

\newcounter{codeLineCntr}

\reversemarginpar
\newif\ifnotes
\notestrue

\newcommand{\punt}[1]{}

\renewcommand{\eqref}[1]{Equation~(\ref{eq:#1})}

\newcommand{\proc}[1]{\ifmmode\mbox{\textsc{#1}}\else\textsc{#1}\fi}

  \newcommand{\func}[1]{\ifmmode\mathrm{#1}\else\textrm{#1}fi} %

\newcounter{remark}[section]

\usepackage{listings}
\usepackage{xcolor}

\usepackage[htt]{hyphenat}
\usepackage{here}

\definecolor{codegreen}{rgb}{0,0.6,0}
\definecolor{codegray}{rgb}{0.5,0.5,0.5}
\definecolor{codepurple}{rgb}{0.58,0,0.82}
\definecolor{backcolour}{rgb}{0.97,0.97,0.95}

\lstdefinestyle{mystyle}{
    backgroundcolor=\color{backcolour},
    commentstyle=\color{codegreen},
    keywordstyle=\color{magenta},
    numberstyle=\tiny\color{codegray},
    stringstyle=\color{codepurple},
    basicstyle=\ttfamily\footnotesize,
    breakatwhitespace=false,
    breaklines=true,
    captionpos=b,
    keepspaces=true,
    numbers=left,
    numbersep=5pt,
    showspaces=false,
    showstringspaces=false,
    showtabs=false,
    tabsize=2
}
\newcommand{\systemname}{\textsf{Funky}}

\usepackage{algorithmic}
\usepackage[ruled,vlined,linesnumbered]{algorithm2e}

\usepackage{caption}
\usepackage{subcaption}

\usepackage{filecontents}

\setlist{noitemsep,topsep=0pt,parsep=0pt,partopsep=0pt}

\usepackage[subtle]{savetrees}

\usepackage{setspace}
\usepackage[margin=5pt,font={stretch=0.9}]{caption}

\begin{document}

\title{\systemname{}: Cloud-Native FPGA Virtualization and Orchestration}

\author{Atsushi Koshiba}
\affiliation{%
  \institution{Technical University of Munich}
  \city{Munich}
  \country{Germany}
}
\email{atsushi.koshiba@tum.de}
\author{Charalampos Mainas}
\affiliation{%
  \institution{Technical University of Munich}
  \city{Munich}
  \country{Germany}
}
\email{charalampos.mainas@tum.de}
\author{Pramod Bhatotia}
\affiliation{%
  \institution{Technical University of Munich}
  \city{Munich}
  \country{Germany}
}
\email{pramod.bhatotia@tum.de}

\renewcommand{\shortauthors}{Atsushi Koshiba et al.}

\begin{abstract}
The adoption of FPGAs in cloud-native environments is facing impediments due to FPGA limitations and CPU-oriented design of orchestrators, as they lack virtualization, isolation, and preemption support for FPGAs. 
Consequently, cloud providers offer {\em no} orchestration services for FPGAs, leading to low scalability, flexibility, and resiliency. 

This paper presents \systemname{}, a full-stack FPGA-aware orchestration engine for cloud-native applications. Funky offers primary orchestration services for FPGA workloads to achieve high performance, utilization, scalability, and fault tolerance, accomplished by three contributions: (1) FPGA virtualization for lightweight sandboxes, (2) FPGA state management enabling task preemption and checkpointing, and (3) FPGA-aware orchestration components following the industry-standard CRI/OCI specifications. 

We implement and evaluate \systemname{} using four x86 servers with Alveo U50 FPGA cards. Our evaluation highlights that \systemname{} allows us to port 23 OpenCL applications from the Xilinx Vitis and Rosetta benchmark suites by modifying 3.4\% of the source code while keeping the OCI image sizes 28.7$\times$ smaller than AMD's FPGA-accessible Docker containers. In addition, \systemname{} incurs only 7.4\% performance overheads compared to native execution, while providing virtualization support with strong hypervisor-enforced isolation and cloud-native orchestration for a set of distributed FPGAs.
Lastly, we evaluate \systemname{}'s orchestration services in a large-scale cluster using Google production traces, showing its scalability, fault tolerance, and scheduling efficiency.
\end{abstract}



\maketitle

\section{Introduction}
\label{sec:intro}

Cloud-native architectures are a promising trend in the cloud, where various workloads designed as collections of small services, i.e., microservices~\cite{ms-arch,cloud-native-app}, are deployed across server nodes by an orchestrator (e.g., Kubernetes~\cite{kubernetes}, Mesos~\cite{apache-mesos}, Swarm~\cite{docker-swarm}, YARN~\cite{hadoop-yarn}) on behalf of users.  
The orchestrators play an important role in efficient cloud resource management, including resource provisioning, workload scaling~\cite{autoscaling-cloud-tsc2024}, scheduling~\cite{borg,borg-2nd,kube-knots,203163}, migration~\cite{state-management-cloud-native,container-live-mig-2020,mwarp,sledge}, and fault tolerance~\cite{9903004,8854724,10419298}. 

In the meantime, hardware accelerators are rapidly adopted by major cloud providers, such as GPUs~\cite{google-cloud-gpu,nvidia-minstance-gpu}, TPUs~\cite{tpu-cloud}, and FPGAs~\cite{ec2f2,alibaba,catapult}, to meet the computational demands of modern cloud workloads. Despite their promising performance benefits, existing cloud orchestrators are natively designed to manage {\em only} CPU and memory resources~\cite{cloud-native-survey2024} and lack comprehensive support for these accelerators. 
Because accelerators are more limited and expensive than CPUs due to a small number of PCIe slots per server node~\cite{ec2f2,azure-np,nvidia-minstance-gpu}, orchestration support for accelerators is crucial to efficiently share them across millions of services running in a multi-tenant cloud. 
 
Although various hardware accelerators are available in modern cloud environments, we primarily focus on FPGAs because of their flexibility and customizability. FPGAs are programmable hardware where users can configure custom logic specialized to a specific computation, making them attractive for ever-changing cloud workloads. As a result, FPGAs are widely offered by all major cloud providers ~\cite{ec2f2,alibaba,catapult}, and are shown to be effective for machine learning~\cite{opencl-cnn,fpga-dcnn}, databases~\cite{10.1145/2554688.2554787}, distributed applications~\cite{tnic}, storage stack~\cite{fvm}, and graph processing~\cite{fpgp,graphops,cygraph}.

Despite FPGAs' promise to accelerate various cloud workloads in an energy-efficient manner,  there is currently {\em no orchestration engine} for FPGAs~\cite{8533482,quraishi2021survey,cloud-native-survey2024}. 
We identify three key challenges that hinder FPGA orchestration, which are not fully addressed by existing industry practices~\cite{xilinx-k8s,intel-fpga-k8s} and academic solutions~\cite{blastfunction,molecule}. 

{\em \textbf{First, no lightweight sandboxes for FPGA applications~exist.}}
An execution sandbox (e.g., VM) is paramount to virtualizing and sharing cloud resources across multi-tenant applications. Cloud-native environments adopt lightweight sandboxes such as containers~\cite{gvisor,cntr} and lightweight VMs~\cite{firecracker,lightvm,vmsh,wallet-arxiv} for small performance penalties~\cite{x-containers} and low start-up latencies~\cite{lupine,catalyzer}. 
However, FPGA virtualization on these sandboxes has not yet been explored. While FPGA vendors offer pre-built Docker containers~\cite{xilinx-base-runtime}, they not only lack FPGA virtualization but also need a large portion of FPGA system stacks installed, sacrificing their lightweight nature. 

{\em \textbf{Second, FPGA-accelerated workloads are not preemptible/recoverable.}} 
Cloud-native environments deploy not only stateless workloads (e.g., serverless functions), to which prior studies~\cite{molecule,blastfunction} adapt FPGAs, but also long-running stateful microservices~\cite{cncf-survey-2020,dok-report-2021,sigmaos} that keep alive and repeatedly handle incoming requests. Stateful workloads are also primary targets for FPGA acceleration, such~as databases~\cite{fpga-memory-db-survey,fpga-nrdb-survey,7577329}, machine learning~\cite{pipecnn,fpga-dl-review}, and search engines~\cite{catapult}. 
However,  the lack of FPGA state management (e.g., context switch) complicates FPGA sharing across them. For instance, once such workloads get and initialize any FPGA, they want to keep it `warmed up' to avoid paying high bootup costs again (e.g., FPGA reconfiguration, input data initialization) for low latencies~\cite{kernel-as-a-service}. 
Allowing long-running services to occupy FPGAs without preemption leads to severe resource underutilization and also data loss by system failures~\cite{9903004}. 

{\em \textbf{Third, existing orchestrators are unaware of FPGA resources.}} 
Existing cloud orchestrators offer limited FPGA support due to the lack of virtualization mechanisms. Device plugins~\cite{xilinx-k8s,intel-fpga-k8s} do not support FPGA sharing across multi-tenant containers. 
Extending orchestrators for FPGA support is challenging because they follow open specifications such as Container Runtime Interface (CRI)~\cite{cri} and Open Container Initiative (OCI)~\cite{oci}, which are not designed for accelerator management. Any extension must not violate these specifications to guarantee system compatibility.

To overcome them, we propose \systemname{}, an FPGA orchestration engine for cloud-native applications. \systemname{} is the first system that tackles an end-to-end orchestration support for cloud FPGAs, achieving three contributions: 
\begin{enumerate}[leftmargin=6mm]
    \item {\bf Lightweight FPGA virtualization}~($\S$~\ref{design:virtualization}, $\S$~\ref{design:opencl}). 
    We design a unikernel-based lightweight sandbox for FPGA virtualization. 
    Unikernels~\cite{10.1145/2490301.2451167,gramine-tdx,uio,unikraft,lightvm} are specialized OSes designed to execute a single, specific application. 
    Our unikernel is designed for FPGA applications, which encapsulates individual CPU applications using FPGAs and isolates them from underlying FPGA system stacks~\cite{vitis-platform,intel-afu,coyote}. It is not only suitable for cloud-native applications but also allows the orchestrator to transparently allocate/deallocate FPGAs to tasks on demand. 
    
    \item {\bf FPGA state management}~($\S$~\ref{design:migration}). We design a task state management mechanism to suspend and resume FPGA workloads. It enables evicting, migrating, and checkpointing tasks whose states are distributed across CPU and FPGA devices. Unikernel's simplified state and single-process basis facilitate the underlying hypervisor in tracing FPGA I/O requests from each guest sandbox and saving/loading CPU and FPGA states.
    
    \item {\bf FPGA-aware orchestration}~($\S$~\ref{design:scheduler}). We demonstrate an end-to-end design integration of \systemname{}'s virtualization and state management mechanisms into industry-standard orchestrators without violating the CRI/OCI specifications~\cite{oci,cri}. It provides three primary FPGA orchestration services: preemptive scheduling, checkpointing, and workload scaling.
\end{enumerate}

We implement the \systemname{} framework for Alveo U50 FPGA with the Vitis Shell~\cite{vitis-platform}. 
We extend IncludeOS~\cite{includeos} and Solo5~\cite{solo5} to implement the unikernel sandbox and dedicated hypervisor. We also implement a prototype of our orchestrator and container runtime, offering high-level orchestration services for FPGA workloads in cooperation with the sandbox. 

We evaluate  \systemname{} across two dimensions. First, we evaluate \systemname{}'s virtualization overheads, portability, and state management using two FPGA benchmark suites, i.e., Vitis benchmarks~\cite{vitis-bench} and Rosetta's real-world applications~\cite{rosetta}, on a real four-node FPGA cluster.
The results highlight that \systemname{} allows us to port 23 OpenCL applications by modifying $3.4\%$ of the source code with $28.7\times$ smaller image size on average than Docker containers maintained by an FPGA vendor (AMD)~\cite{xilinx-base-runtime}. \systemname{} imposes performance overheads of $7.4\%$ compared to the native execution, only $0.6\%$ higher than AMD's containers that do not virtualize FPGAs.
Second, we evaluate \systemname{} using Google production traces~\cite{borg-2nd} to showcase its orchestration effectiveness for a large-scale cluster. 

\begin{figure*}[t]
  \begin{center}
    \includegraphics[width=0.95\linewidth]{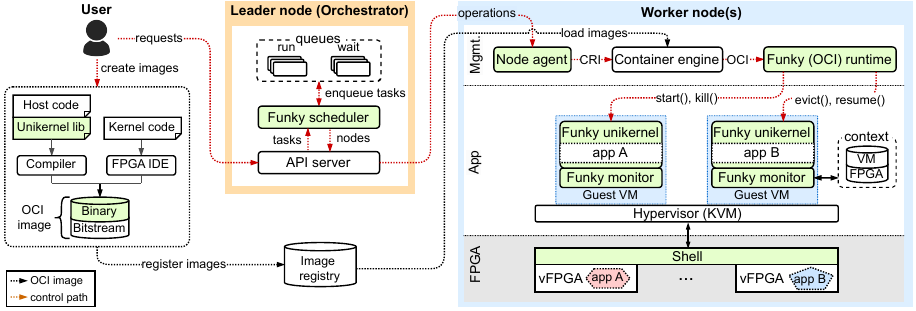}
  \end{center}
  \caption{The \systemname{} architecture for cloud-native FPGA orchestration. The key components are highlighted as green boxes.}
  \label{fig:k8s-overview}
   \vspace{-4mm}
\end{figure*}

\section{Background and Motivation}
\subsection{FPGAs in Cloud Environments}

This paper targets cloud-native environments, where applications consist of multiple small tasks (e.g., microservices) that are scheduled and deployed by a cloud orchestrator (e.g., Kubernetes~\cite{kubernetes}) across distributed nodes. Deployed tasks running on CPUs are isolated by guest sandboxes (VMs~\cite{firecracker,lightvm}, containers~\cite{lxc,gvisor}) and can be dynamically evicted, resumed, and migrated by the orchestrator for load balancing and auto-scaling.
Each task is built as an OCI image~\cite{oci-image} containing its executable and guest sandbox image. 

\myparagraph{FPGA architecture}
In cloud infrastructures, FPGAs serve as standalone PCIe devices connected to each machine node~\cite{ec2f2,alibaba,azure-np}. 
The device contains FPGA fabric and onboard peripherals such as DDR memory, network ports, and a PCIe slot.
A portion of the FPGA fabric is statically configured as a \textit{Shell}~\cite{vitis-platform,intel-afu}, which offers glue logic to connect to the onboard peripherals, such as a PCIe bridge, DMA, and network controller. The rest serves as a dynamic region for runtime custom logic reconfiguration.

\myparagraph{FPGA applications}
Like other accelerators, FPGAs are mainly used as coprocessors, where CPU applications offload their compute-intensive part and receive results after completion. Such FPGA applications consist of two parts: the offloaded computation part called \textit{kernel code} and the main CPU application called \textit{host code}. 

The kernel code represents custom logic (or \textit{user logic}) that executes the offloaded task. 
The kernel code can be written in any language supported by the underlying FPGAs (e.g., HDL~\cite{verilog,vhdl}, HLS~\cite{rsass,lime,opencl,spatial,vsipl}, and DSLs~\cite{chisel,linqits,7551387,10.1155/2012/439141}). 
The code is compiled using FPGA vendors' IDEs (e.g., Vivado~\cite{vivado}), which generate \textit{bitstreams}. 
The bitstream is programmed to the FPGA's dynamic region to instantiate user logic there. 

The host code is responsible for FPGA device management (e.g., reconfiguration) and task offloading via various vendor-provided APIs~\cite{oneapi,opencl,xrtapi}.
We primarily target the OpenCL API~\cite{opencl} because it is widely adopted by FPGA vendors~\cite{vitis-platform,intel-opencl}, and other APIs (e.g., XRT (\cite{xrtapi}), OneAPI (\cite{oneapi})) inherit its programming model.

\subsection{Design Challenges and Key Ideas}
\label{subsec:challenges}
FPGA adoption in cloud-native ecosystems has not been fully explored by existing studies targeting traditional cloud instances~\cite{optimus,ava} or serverless platforms~\cite{molecule,blastfunction}.
We identify three key features to realize successful FPGA integration into cloud-native architectures, filling a gap in prior studies. 
First, \emph{a lightweight FPGA virtualization mechanism} is essential, which abstracts physical FPGA devices from guest cloud-native applications (host code) running inside CPU sandboxes without compromising their lightweightness. This feature facilitates sharing limited FPGA resources among multi-tenant applications and also enables the orchestrator to transparently allocate/deallocate FPGA resources. 
Second, \emph{an FPGA state management mechanism} is critical for core orchestration operations such as preemptive scheduling, checkpointing, and load balancing. This feature must allow the orchestrator to safely save and restore application contexts residing in both CPU and FPGA devices. 
Lastly, we need \emph{an end-to-end FPGA orchestration mechanism} that can be integrated into industry-standard orchestrators such as Kubernetes. This demonstrates the compatibility and applicability of our solutions within modern cloud-native ecosystems. 

The FPGA-aware orchestration ecosystem must overcome design challenges related to FPGA constraints and prior solutions' limitations. We present four core challenges and \systemname{} approach.


\myparagraph{\#1: Lightweight FPGA virtualization}
Existing lightweight sandboxes~\cite{gvisor,firecracker,lightvm,unikraft} only support virtualization for legacy devices such as disk and network, and FPGA support is missing.  
Prior studies propose FPGA virtualization for traditional cloud instances~\cite{ava,optimus}, which are impractical for cloud-native environments due to a lack of reconfiguration support~\cite{optimus} or the complexity of the guest sandbox/API layer~\cite{ava}. 
While rich sandboxes ensure system compatibility, their layered stacks obscure the interaction between guest applications and FPGAs, complicating orchestration services depending on their states, e.g., preemption. 

\myidea{} We design a new unikernel architecture virtualizing FPGAs for cloud-native applications, ensuring the lightweight context, near-zero overhead, and hypervisor-enforced isolation (see $\S$~\ref{design:virtualization}). 

\myparagraph{\#2: Application portability}
Packaging FPGA applications as unikernel images imposes another challenge because of the lack of programmability. To address this issue, we aim to support a modern programming API for FPGAs, i.e., OpenCL, to retain application portability. Standard VMs or containers are capable of porting the entire FPGA software stacks to virtualized environments~\cite{optimus,xilinx-k8s}. However, porting the vendor OpenCL stack to the unikernel enlarges application contexts and undermines its lightweight architecture. 

\myidea{} We introduce FunkyCL, a lightweight OpenCL-compatible library that gains programmability for OpenCL execution workflows with minimal code changes (see $\S$~~\ref{design:opencl}).

\myparagraph{\#3: FPGA task preemption and state management}
Although cloud service preemption, migration, and checkpointing are active research areas~\cite{RePEc:wly:intnem:v:32:y:2022:i:6:n:e2212,kaur:hal-03466765,SINGH2024101887,deng2023cloudnative,9903004,MManager,8854724,10419298}, they do not take accelerator device states (FPGAs, GPUs) into account. Prior studies propose hardware-assisted checkpoints for FPGA preemption~\cite{10.5555/647927.739382,7284295,Knodel}, which integrate monitoring circuits into user logic to capture the FPGA's internal states. However, saving only the logic states is insufficient because the context of FPGA applications is distributed across CPUs, FPGAs, and their memories. This state distribution and FPGA's asynchronous execution model~\cite{opencl,xrtapi} complicates orchestrators' ability to maintain the entire state of FPGA workloads.  

\myidea{} We propose an end-to-end state management mechanism for FPGA task preemption and checkpointing. The thin hypervisor transparently maintains both VM and FPGA contexts (see $\S$~\ref{design:migration}).

\myparagraph{\#4: FPGA-aware orchestration}
Integrating \systemname{}'s FPGA virtualization and preemption mechanisms into cloud-native orchestrators poses two design challenges. First, industry-standard orchestrators, e.g., Kubernetes, do not support checkpoint and restore operations even without FPGAs. Although the Kubernetes scheduler supports preemption and eviction for a group of containers (i.e., pods)~\cite{k8s-eviction}, these functions terminate the containers and abandon their contexts. Second, the preemption requests must be propagated from the orchestrator to worker-node components while following the CRI/OCI specifications~\cite{cri,oci} for system compatibility. 

\myidea{} We design CRI/OCI-compatible orchestration system components that leverage the proposed FPGA virtualization and state management mechanisms (see $\S$~\ref{design:scheduler}). 

\section{Design}
\label{overview}

\subsection{System Overview}

\myparagraph{Cloud-native architecture} 
Figure~\ref{fig:k8s-overview} illustrates the \systemname{}'s cloud-native ecosystem. We target an industry-standard cloud-native system stack following the CRI and OCI specifications~\cite{cri,oci}.
The cloud orchestrator runs on the leader node, whereas per-node system components and user applications run on the worker nodes. Every worker node is equipped with one or more PCIe-connected FPGAs, and each FPGA offers a dedicated Shell, onboard memory, and one or more reconfigurable slots (vFPGAs) where user logic is programmed. 
The orchestrator is responsible for deploying applications, managing the entire cluster, and enforcing scheduling policies. The cloud users push OCI images of applications to the image registry before sending deployment requests to the orchestrator. The node agent runs on every worker node and forwards requests from the orchestrator to the container engine through CRI APIs. Upon API calls from the agent, the engine invokes the corresponding commands of the OCI runtime. 

We highlight three system components extended/added to realize FPGA-aware orchestration. \emph{\systemname{} scheduler} is the extended scheduler component, which selects a worker node appropriate for deploying/migrating applications based on the applied policy. 
\emph{\systemname{} runtime} is an OCI-compatible runtime responsible for deploying, evicting, resuming, and migrating applications in response to orchestration requests. 
\emph{\systemname{} monitor} is a thin hypervisor layer that virtualizes FPGAs on worker nodes from unikernel applications. 

\myparagraph{Build process and tool chain}
Figure~\ref{fig:k8s-overview} also illustrates a workflow of packaging applications as deployable images on the \systemname{} architecture. Cloud users prepare an application binary and corresponding FPGA bitstreams. The binary is generated by compiling the host code statically linked with the \systemname{} unikernel library, which exposes guest APIs for FPGA control to applications.
The bitstream is generated from the kernel code with FPGA development tools such as Vivado~\cite{vivado}. The kernel code does not change because \systemname{} runs upon the original vendor-provided FPGA platform~\cite{vitis-platform}.

\label{design}

\begin{table*}[t]
    \centering
    \rowcolors{2}{}{lightgray!40}
    \begin{tabular}{|p{43mm}|p{66mm}|p{57mm}|} \hline
        {\bf OpenCL APIs} & {\bf Standard OpenCL definition} & {\bf FunkyCL definition} \\ \hline
        clCreateProgramWithBinary()  & Creates a program object and loads bitstreams. & Calls \texttt{vfpga\_init()} to configure user logic. \\ 
        clReleaseProgram()           & Decrements the program reference count. & Calls \texttt{vfpga\_free()} if the count gets zero. \\ 
        clCreateBuffer()             & Creates an OpenCL buffer (memory) object.        & Sends a  {MEMORY()} request. \\ 
        clEnqueueKernel()            & Enqueues a command to execute a kernel.          & Sends an {EXECUTE()} request. \\ 
        clEnqueueMigrateMemObjects() & Enqueues a command to migrate memory objects.    & Sends a  {TRANSFER()} request. \\ 
        clFinish()                   & Waits until all enqueued commands are complete.  & Sends a  {SYNC()} request. \\ \hline
    \end{tabular}
    \caption{Definitions and functional changes of important OpenCL host APIs~\cite{opencl}; other APIs are omitted for space.}
    \vspace{-3mm}
    \label{tab:opencl_apis}
\end{table*}

\subsection{Lightweight FPGA Virtualization}
\label{design:virtualization}
We design the \systemname{} unikernel, a lightweight, isolated guest sandbox virtualizing distributed FPGAs. The \systemname{} unikernel virtualizes distributed FPGAs in an orchestrator-friendly manner. 
First, the lightweight state of a unikernel mitigates start-up latencies and virtualization overheads. Second, its simplified state and single-process basis facilitate FPGA task preemption and migration across distributed nodes. The underlying hypervisor can easily trace memory transfers and operations among the guest application and FPGAs.

The \systemname{} unikernel delegates most FPGA-related jobs requested by applications to a host-side hypervisor layer. 
This approach is similar to API remoting~\cite{vocl,ava}. However, the \systemname{} unikernel does not directly delegate APIs but converts them to minimal I/O requests essential for FPGA task offloading (e.g., data copy and synchronization) to securely isolate guest applications and the host system stack. 

\systemname{}'s FPGA virtualization mechanism is technically applicable to other hypervisor-based sandboxes, e.g., standard full-blown VMs and Firecracker's microVMs~\cite{firecracker}. \systemname{} adopts a unikernel as a guest sandbox because of increasing security concerns in cloud-native environments, given that even containers are running inside VMs (e.g., Kata container~\cite{kata-container}). We strive to design the unikernel-based virtualization mechanism without losing the security properties of hypervisor-enforced isolation.

\begin{figure}[t]
\begin{center}
  \includegraphics[width=0.95\linewidth]{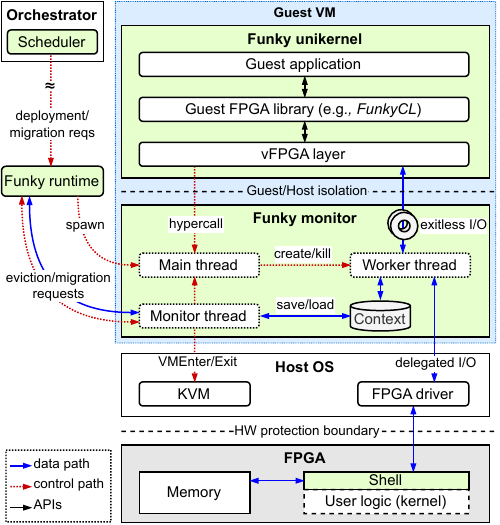}
\end{center}
  \vspace{-1mm}
  \caption{\systemname{}'s unikernel architecture.}
  \vspace{-2mm}
  \label{fig:unikernel}
\end{figure}

\myparagraph{Unikernel design} 
Figure~\ref{fig:unikernel} shows the \systemname{} unikernel architecture. 
\textit{The \systemname{} unikernel} virtualizes reconfigurable FPGA slots as \textit{vFPGAs} and allows applications to request managing vFPGAs through guest APIs. 
The unikernel exposes \textit{vFPGA layer} to a guest application and library, virtualizing the host FPGA and offering abstracted interfaces: hypercalls and exitless I/O~\cite{io-uring,rkt-io}. The guest FPGA library converts API calls from the guest application to either the hypercalls or exitless I/O requests. In this work, we design \textit{FunkyCL}, a lightweight OpenCL-compatible guest library ($\S$~\ref{design:opencl}). 

\textit{The \systemname{} monitor} is a host-side user process serving as a thin hypervisor layer for each \systemname{} unikernel, similar to QEMU~\cite{qemu}. The \systemname{} monitor is responsible for launching the corresponding unikernel and monitoring its states. 
It is also responsible for handling requests from the guest application and the orchestrator while ensuring host-side isolation among multiple \systemname{} monitors.
To handle these asynchronous requests efficiently, the monitor creates two threads: a \textit{worker thread} and \textit{monitor thread}. The former handles FPGA control requests from the guest and manages the underlying FPGA via a host FPGA system stack (e.g., Xilinx Runtime~\cite{xrtapi}). The latter handles migration/eviction requests from the orchestrator and triggers saving/loading guest VM and FPGA contexts.

\myparagraph{Hypercalls}
The \systemname{} unikernel offers only two hypercalls used for vFPGA allocation. {vfpga\_init()} requests the \systemname{} monitor to search for an available vFPGA and assigns it to the guest application. The hypercall also transfers a bitstream to let the \systemname{} monitor reconfigure the vFPGA. Subsequently, the \systemname{} monitor creates the worker thread to asynchronously handle requests from the guest. The requests are forwarded via lock-free message queues shared between the unikernel and the monitor. The \systemname{} unikernel initializes the queues in the guest memory space and notifies their addresses to the monitor via the hypercall. Conversely, {vfpga\_exit()} releases the obtained vFPGA and deletes the worker thread.

\myparagraph{Asynchronous I/Os}
The \systemname{} unikernel offers a fast, exitless I/O interface for FPGAs to avoid frequent context switches.
To realize this, we design \systemname{} requests, FPGA-related primitive operations to be handled on the underlying FPGA platform. Table~\ref{tab:funky_request} shows four primary \systemname{} requests. A guest unikernel sends the \systemname{} requests to the worker thread via shared queues without invoking {VMEXIT}. The worker thread securely validates the received \systemname{} requests and performs the delegated I/Os upon the requests. 
We note that the \systemname{} requests listed in Table~\ref{tab:funky_request} are designed for OpenCL support ($\S$~\ref{design:opencl}). 
Other \systemname{} requests can be further designed to support different programming models. 

\subsection{FunkyCL Library}
\label{design:opencl}
We design FunkyCL, an OpenCL-compatible library that brings application portability and compatibility for \systemname{} unikernel applications.
FunkyCL is designed to realize two key functionalities. First, it is fully compatible with the native OpenCL specification standardized by Khronos group~\cite{opencl}. Second, FunkyCL APIs achieve performance equivalent to the native execution of OpenCL APIs.

\begin{table}[]
    \setlength\tabcolsep{3.6pt}
    \centering
    \rowcolors{2}{}{lightgray!40}
    \begin{tabular}{|p{40mm}|p{40mm}|} \hline
        {\bf Request type} & {\bf Description} \\ \hline
        {MEMORY(}\textit{buff\_id, src, size}{)}          & Allocates a buffer on FPGA.  \\ 
        {TRANSFER(}\textit{queue, buff\_id, src, size}{)} & Invokes a data transfer between Host \& FPGA memories.  \\ 
        {EXECUTE(}\textit{queue, kernel, args}{)}        & Invokes a kernel execution.  \\ 
        {SYNC(}\textit{queue, req\_id}{)}                & Awaits request completion.  \\ \hline
    \end{tabular}
    \caption{\systemname{} requests for FPGA operations. }
    \vspace{-2mm}
    \label{tab:funky_request}
\end{table}

\myparagraph{OpenCL APIs}
The FunkyCL library is part of the \systemname{} unikernel library and allows the guest application to control the assigned vFPGA via the standard OpenCL APIs. The API calls issue the \systemname{} requests described in Table~\ref{tab:funky_request}.
FunkyCL offers a library-level FPGA abstraction to enable transparent use of the underlying FPGAs; it exposes the \systemname{} unikernel's vFPGA layer as an OpenCL device named \textit{vFPGA}. OpenCL host APIs issued to the device are transparently converted to either hypercalls or \systemname{} requests. 

Table~\ref{tab:opencl_apis} lists important OpenCL APIs and functional changes in FunkyCL.
{clCreateProgramWithBinary()} and {clReleaseProgram()} play an essential role: acquiring and releasing vFPGAs. 
Because creating a program object triggers FPGA reconfiguration, FunkyCL lets {clCreateProgramWithBinary()} invoke {vfpga\_init()} to obtain the vFPGA. The \systemname{} monitor handles the hypercall and spawns the worker thread if it finds an available vFPGA. Subsequently, the worker thread reconfigures the slot with a bitstream. The other OpenCL APIs in Table~\ref{tab:opencl_apis} send corresponding requests to the worker thread until the vFPGA is released~by~{clReleaseProgram()}.

\myparagraph{Zero-copy host-device data transfers}
As a guest application runs in an isolated context in \systemname{}, the host-device data transfer is challenging; such data transfers typically induce redundant data copies, such as shadow paging~\cite{10.1145/1168857.1168860}, to keep data consistent. 
FunkyCL avoids the memory management overhead thanks to the unikernel's single address space. Specifically, when a guest application calls any OpenCL API for data transfers, FunkyCL creates a {TRANSFER()} request, which only contains the address and data size of the guest memory buffer. The worker thread receives the request and then translates the guest address to the host only once. Finally, it transfers the buffer data to the FPGA memory or vice versa.

\subsection{FPGA State Management}
\label{design:migration}
We design an FPGA state management mechanism, enabling task preemption/migration and checkpointing to improve scheduling fairness and fault tolerance. To maintain application contexts distributed across CPUs and FPGAs, \systemname{} adopts a hypervisor-driven approach, i.e., driven by the \systemname{} monitor. 
Since the \systemname{} monitor is spawned per guest VM and handles delegated FPGA requests, it can easily trace both guest VM (CPU) and FPGA states.

\systemname{} supports three state management commands: \textit{checkpoint} that saves snapshots of the entire VM/FPGA contexts, \textit{evict} that only evicts the FPGA context in host memory, and \textit{resume} that resumes the task from either the saved snapshot or evicted context. These commands are triggered by an extended container runtime, i.e., Funky runtime ($\S$\ref{design:scheduler}), in response to high-level orchestration operations requested by the orchestrator. To enable this, the \systemname{} monitor spawns a \textit{monitor thread} that exposes an inter-process communication (IPC) interface to let the \systemname{} monitor issue these commands.  
The monitor thread is also responsible for saving and restoring the VM context and cooperating with the worker thread to maintain an FPGA context.

\begin{figure}[t]
\begin{center}
  \includegraphics[width=0.95\linewidth]{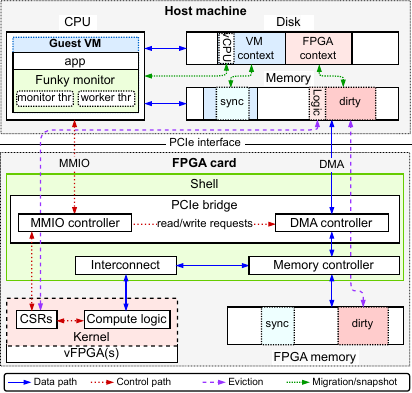}
\end{center}
  \caption{\systemname{}'s hardware architecture.}
  \vspace{-2mm}
  \label{fig:shell}
\end{figure}

\myparagraph{FPGA architecture}
We first describe our target FPGA architecture and guest application states, illustrated in Figure~\ref{fig:shell}. \systemname{} primarily targets vendor-provided Shells~\cite{alveo-platforms,intel-ofs,intel-opencl}. 
The FPGA device consists of three hardware components: Shell, vFPGA(s), and off-chip FPGA memory. The Shell offers the PCIe bridge IP that exposes memory-mapped I/Os to host CPUs and the DMA controller for data transfers between the host and FPGA memory. The vFPGAs configure user logic, which could have control and status registers (CSRs) for the execution invocation. The FPGA memory (HBM, DRAM) and its controller are used as the main working memory, where we store input/output data consumed/produced by the kernel. 
In the architecture, the states of FPGA applications can be classified into three types: the FPGA logic (kernel), FPGA memory, and the guest VM.

\myparagraph{FPGA synchronization}
We next describe how these application states are saved by \systemname{}. 
Because FPGAs do not support context switch mechanisms due to their complex and modular architectures, we cannot suspend any operations running on the FPGA, e.g., DMA transfers and kernel executions. Therefore, before saving the FPGA states, \systemname{} tolerates waiting for all on-the-fly FPGA operations (i.e., \systemname{} requests) to complete. Specifically, when the monitor thread receives checkpointing or eviction requests, it asks the worker thread to invoke the SYNC() request and makes the FPGA state (both user logic and Shell) \textit{consistent}. 
We note that such synchronization operations do not increase the total execution time of running applications because any computation requested by the application is still running during the synchronization. 

While the synchronization operations do not incur performance overheads, they may delay invoking eviction and migration requests and affect the latency of orchestration operations, particularly for a single, long-running request, e.g., processing a 1~GiB chunk. 
Therefore, \systemname{} supports a mechanism to mitigate the synchronization time by splitting such a large request into multiple requests for smaller chunks, particularly for streaming computation that repeats the same operation for all inputs (e.g., FFT). 

\myparagraph{FPGA logic state} Following synchronization, the worker thread saves the FPGA's kernel and memory states. The logic state comprises CPU-readable control registers and temporal contents stored in on-chip elements (flip-flops, BRAM). 
While the latter is not accessible from the host, these internal states are ignorable for most of the state-of-the-art FPGA accelerators following the OpenCL execution model~\cite{opencl-cnn,opencl-cnn2,7577329,7859319,pipecnn,hls4ml,sql2fpga} because their internal states are flushed when the execution completes.  
\systemname{} can technically support saving/restoring FPGA internal states by leveraging hardware-assisted FPGA checkpointing~\cite{synergy,statereveal,10.5555/647927.739382,7284295}. 

\myparagraph{FPGA memory state}
\systemname{} efficiently saves and restores FPGA memory contexts depending on the states of each memory buffer, via DMA data transfers. 
During the execution, the worker thread traces {MEMORY()} requests and reserves locations of memory buffers. Then, upon {TRANSFER()} request, it updates their states as follows: 
\begin{itemize}
  \setlength{\parskip}{0cm} 
  \setlength{\itemsep}{0cm} 
  \setlength{\itemindent}{-4mm}
  \item \textit{init} -- the buffer has no data in the FPGA memory. 
  \item \textit{sync} -- the buffer is synchronized with a host buffer.
  \item \textit{dirty} -- the buffer is not synchronized with a host buffer.
\end{itemize}
When saving FPGA contexts, \systemname{} only records the states of \textit{dirty} memory buffers. Memory buffers whose state is either \textit{init} or \textit{sync} are ignored to mitigate the total context size. The monitor thread kills the worker thread after all the FPGA states are saved. 

\myparagraph{VM state}
After saving FPGA contexts, the monitor thread optionally saves VM contexts. \systemname{} adopts a simple hypervisor-driven approach for saving and restoring VM states. When saving the state as a snapshot, the monitor thread first triggers an interrupt to the vCPU, causing VMEXIT. It then captures the context of the vCPU (e.g., CPU registers) and dirty pages in guest memory. 
When restoring the state, the thread initializes the vCPU and guest memory contexts with the selected snapshot and lets the vCPU invoke VMENTER. The eviction process does not save the VM state to the disk but keeps it in the host memory to mitigate the overhead. 

\myparagraph{Restoring states}
There are two ways to restore the application state from saved contexts and resume the execution. 
First, if only the FPGA states are evicted, the monitor thread simply respawns the worker thread with the saved context and lets it restore the register and memory states of FPGA kernels via memory-mapped I/O and DMA transfers. The worker thread then notifies the completion of a {SYNC()} request to the guest application to resume the execution. 
Second, if the application is resumed from the entire snapshot, the \systemname{} monitor is spawned with the snapshot and first restores the VM state in CPU memory, which includes some of the FPGA contexts, e.g., contents of \textit{sync} buffers. Then, it spawns the worker thread and restores FPGA states as explained. 

\myparagraph{Memory management and isolation}
The \systemname{} monitor guarantees FPGA memory isolation across guest VMs for multi-tenancy. The \systemname{} monitor restricts access to vFPGAs and FPGA memory buffers that are allocated to other applications. To mitigate side-channel risks, the \systemname{} monitor zeros the memory buffers when a guest application finishes.

\begin{table*}[t]
    \centering
    \rowcolors{2}{}{lightgray!40}
    \begin{tabular}{|l|l|l|l|} \hline
        \textbf{Orchestration Services} &\textbf{Operations} & \textbf{CRI API (\textit{key metadata})} & \textbf{\systemname{} OCI runtime command} \\ \hline
        \textbf{Preemptive scheduling}  & Deploy        & CreateContainer (\textit{preemptible*}) $\xrightarrow{}$ StartContainer (\textit{cid})   & create <\textit{cid}> $\xrightarrow{}$ start <\textit{cid}>    \\
                                        & Evict         & StopContainer (\textit{cid})               & evict  <\textit{cid}>   \\
                                        & Resume        & StartContainer (\textit{cid})              & resume <\textit{cid}>   \\
                                        & Migrate       & CreateContainer (\textit{cid*, node\_id*}) $\xrightarrow{}$ StartContainer (\textit{cid}) & resume <\textit{cid, node\_id}>   \\ \hline
        \textbf{Checkpoint \& restore} & Checkpoint    & CheckpointContainer (\textit{cid}) & checkpoint <\textit{cid}>\\
                                        & Restore       & CreateContainer (\textit{cid*, node\_id*}) $\xrightarrow{}$ StartContainer (\textit{cid}) & resume <\textit{cid, node\_id}>\\ \hline
        \textbf{Workload scaling}       & Horizontal    & CreateContainer (\textit{cid*, node\_id*}) $\xrightarrow{}$ StartContainer (\textit{cid}) & replicate <\textit{cid, node\_id}>\\
                                        & Vertical      & UpdateContainerResources (\textit{cid, vfpga\_num*}) & update <\textit{cid, vfpga\_num}>   \\ \hline
    \end{tabular}
    \caption{\systemname{}'s orchestration services. \textit{cid} abbreviates \textit{container id}. * represents metadata attached by \textit{annotations}. }
    \vspace{-2mm}
    \label{tab:orchestration_services}
\end{table*}

\subsection{FPGA-aware Orchestration}
\label{design:scheduler}
Lastly, we design an FPGA-aware orchestrator that leverages \systemname{}'s FPGA virtualization and state management mechanisms. 
Assuming that the use of FPGAs is selective depending on workloads, we strive to achieve two design goals. First, our design can adapt existing industry-standard orchestrators (e.g., Kubernetes) by reasonably extending them without violating CRI and OCI specifications. Second, the extended orchestrator can distinguish between FPGA and CPU workloads, applying  \systemname{}-specific features only to FPGA ones.

\myparagraph{Orchestration Components}
To integrate \systemname{} features into industry-standard orchestrators, we extend three system components: the scheduler, node agent, and OCI runtime. 

\emph{The \systemname{} scheduler} is an orchestration component for FPGA task deployment and preemption. It adopts the same task eviction mechanism as the Kubernetes scheduler~\cite{k8s-eviction}, evicting running tasks if they occupy resources (e.g., vFPGAs) requested by high-priority tasks. However, unlike the Kubernetes scheduler that always terminates evicted tasks, \systemname{} keeps the context of evicted tasks and transparently deallocates resources from them. The evicted tasks can be either \textit{resumed} on the same node or \textit{migrated} to another node. 

\emph{Node agents} propagate requests from the orchestrator to the container engine through CRI APIs. 
To forward \systemname{}-specific information, we leverage \textit{annotations}, unstructured key-value pairs defined in the CRI API's message structure. The node agents attach FPGA-specific metadata to annotations of primary CRI APIs, allowing the container engine to invoke corresponding Funky runtime commands without violating the CRI specification.  

\emph{\systemname{} runtime} is an OCI-compliant low-level container runtime that maintains a lifecycle of \systemname{} unikernels. In addition to commands defined in the OCI specification~\cite{oci}, it supports five \systemname{}-specific commands: {evict}, {resume}, {checkpoint}, {replicate}, and {update}.

{
 \newcommand{\myfontsizealgorithms}{\fontsize{9}{9}\selectfont}
 \LinesNumbered
 \setlength{\parskip}{0pt}
 \begin{algorithm}[t]
 \myfontsizealgorithms
 \SetAlgoLined
 \SetKwProg{Fn}{Function}{:}{}
 
 \newcommand\mycommfont[1]{\small\ttfamily\textcolor{codegreen}{#1}}
 \SetCommentSty{mycommfont}
 \SetNoFillComment
 

 
 \underline{{\bf schedule($wait\_queue$, $run\_queue$, $nodes$)}} \\
 \Begin{
    \tcc{Pick the most prioritized task}
    task $\gets$ {pull}(wait\_queue)\; 

    \tcc{Find the most suitable node for the task}
    node $\gets$ {schedule}(task, nodes)\; 

    %

    \tcc{Evict, resume, migrate, or deploy the task}
    \uIf{node is occupied by other tasks}{ 
      evicted\_task $\gets$ {evict\_req}(task, node, run\_queue)\; 
      {push}(evicted\_task, wait\_queue)\;
      {deploy\_req}(task, node)\;
    }
    \uElseIf{task is evicted \textbf{and} task is on node}{
      {resume\_req}(task, node)\; 
    }
    \uElseIf{task is evicted \textbf{and} task is not on node}{
      {migrate\_req}(task, node)\; 
    }
    \uElse{
      {deploy\_req}(task, node)\; 
    }

    {push}(task, run\_queue)\;
 }
 \caption{\systemname{}'s preemptive task scheduling. }
 \label{alg:sched}
 \end{algorithm}
 }

%
%
%

\myparagraph{Orchestration services}
Table~\ref{tab:orchestration_services} summarizes \systemname{}'s primary orchestration services and associated operations. 
To distinguish FPGA and non-FPGA tasks, \systemname{} attaches a \textit{preemptible} flag to deployed FPGA tasks. Any orchestration requests for \textit{preemptible} tasks are propagated to the \systemname{} runtime to invoke \systemname{} operations.
The \textit{node\_id} represents a worker node where the context of the target task exists. When migrating, restoring, or replicating a task, the \systemname{} runtime communicates with the other runtime on a remote node specified by \textit{node\_id} to obtain the task context. 
\textit{vfpga\_num} represents the maximum number of vFPGAs allocatable to a task, which is used for vertical scaling.

\myparagraph{Preemptive task scheduling}
Algorithm~\ref{alg:sched} details the workflow of \systemname{}'s preemptive task scheduling. The \systemname{} scheduler has a \textit{run queue} holding running tasks and a \textit{wait queue} holding submitted/evicted tasks sorted by their priorities. 
The scheduler picks up a task to schedule next from the wait queue and selects the most suitable node to place the task (L3-4). 
The scheduler then \textit{evicts}, \textit{resumes}, \textit{migrates}, or \textit{deploys} the task depending on its state. 
When other running tasks occupy the selected node (L5-8), the scheduler \textit{evicts} any of them; it moves the task from the run queue to the wait queue and sends the eviction request to the node agent. The evicted task state is reserved on the same worker node until resumed or migrated. 
When rescheduling an evicted task, depending on the selected node, the scheduler either \textit{resumes} the task on the same node (L9-10) or \textit{migrates} the task to the remote node (L11-12). 

\myparagraph{Checkpoint and restore}
\systemname{} supports both automatic and manual checkpointing functions. In the former case, the orchestrator periodically takes snapshots of the target task, and if the orchestrator detects task failures, it attempts to restore it from the latest snapshot. 
In the latter case, these operations are manually invoked upon user requests on demand. 
The failed tasks can be restored on either the current (local) node or any remote node. The execution flow is the same as the \textit{evict} and \textit{migrate} of the preemptive scheduling. 

\begin{figure*}[t]
  \begin{center}
    \includegraphics[width=\linewidth]{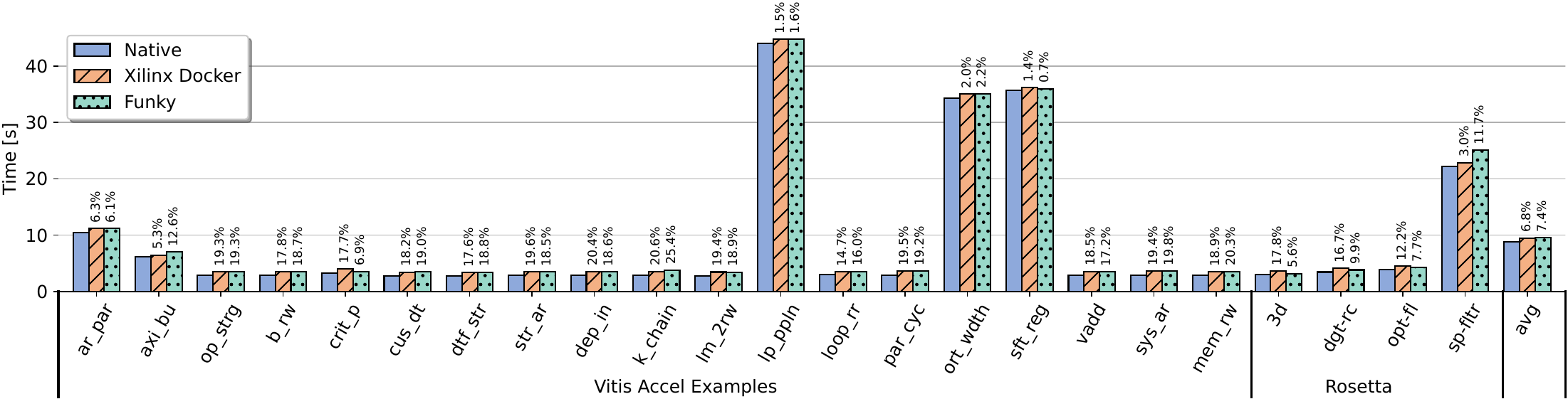}
  \end{center}
  \caption{End-to-end application execution time for native execution, Xilinx Docker container~\cite{xilinx-base-runtime}, and \systemname{}, with relative gaps (\%) to the native execution. 
  }
   \vspace{-2mm}
  \label{fig:e2e_time}
\end{figure*}

\begin{figure*}[t]
    \centering
    \begin{minipage}[b]{0.48\textwidth}
      \centering
      \includegraphics[width=0.9\linewidth]{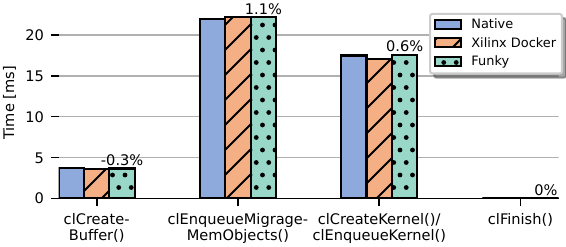}
      \vspace{-1mm}
      \caption{OpenCL API times with relative gaps (\%) to the native execution. 
      }
      \vspace{-2mm}
      \label{fig:oh_funkycl}
    \end{minipage} \hspace{5mm}
    \begin{minipage}[b]{0.48\textwidth}
        \centering
        \includegraphics[width=0.9\linewidth]{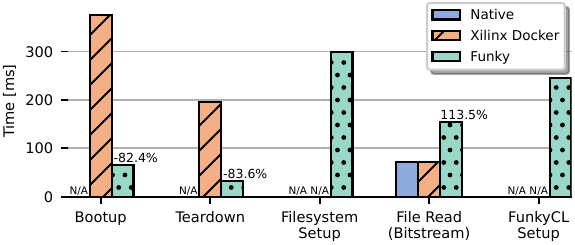}
        \vspace{-1mm}
        \caption{\systemname{}'s setup overheads with relative gaps (\%) to the Xilinx Docker container. 
        }
        \label{fig:overhead_breakdown}
    \end{minipage} \hfill
    \vspace{-1mm}
\end{figure*}

\myparagraph{Workload scaling}
\systemname{} supports both horizontal and vertical scaling for FPGA workloads. 
For the horizontal scaling, \systemname{} \textit{replicates} a running task by saving and copying its context. 
However, if the target task is still alive and running, the container engine invokes \textit{replicate} operation to deploy the replicated task on the node represented by \textit{node\_id}. 
For the vertical scaling, \systemname{} increases the task's limit of allocatable vFPGAs given by \textit{vfpga\_num}.  

\section{Implementation}
\label{sec:implementation}
We build a prototype of the \systemname{} framework targeting the AMD Vitis platform~\cite{vitis-platform} built upon the Alveo U50 FPGA~\cite{alveo-u50} with the XDMA Shell, which serves a single reconfigurable slot.

\myparagraph{\systemname{} unikernel and FunkyCL}
We implement the \systemname{} unikernel and FunkyCL library based on the IncludeOS unikernel~\cite{includeos}. We port open-sourced OpenCL headers~\cite{opencl-header} and C++ bindings~\cite{opencl-cpp} to IncludeOS and implement their entities as a part of the static unikernel library. Some FunkyCL APIs do not issue requests to the backend to mitigate the communication overhead. For example, kernel arguments set by {clSetKernelArg()} are transferred together through the {EXECUTE()} request. 

\myparagraph{\systemname{} monitor}
We implement the \systemname{} monitor based on the Solo5~\cite{solo5} hypervisor, which consists of frontend bindings that offer application binary interfaces (ABI) to the guest unikernel, and a backend hypervisor layer (called ukvm) that acts as a vCPU thread. We mainly extend ukvm to implement the core functionalities, including the FPGA worker thread. The FPGA worker thread handles \systemname{} requests from the guest through APIs offered by Xilinx Runtime (XRT)~\cite{xrtapi}. We also implement two hypercalls, {vfpga\_init()} and {vfpga\_free()}, for obtaining and releasing FPGA slots. 

\myparagraph{Orchestrator and runtime}
Our orchestrator prototype consists of the API server handling orchestration service requests (e.g., Deploy, Checkpoint) and the scheduler for preemptive scheduling. It propagates the respective operations to the \systemname{} runtime daemons on worker nodes. The \systemname{} runtime then issues corresponding commands shown in Table~\ref{tab:orchestration_services}.  

\section{Evaluation}
We comprehensively evaluate \systemname{}'s virtualization overheads ($\S$~\ref{subsec:performance}), portability ($\S$~\ref{subsec:portability}), checkpointing ($\S$~\ref{subsec:checkpointing}), task preemption ($\S$~\ref{subsec:preemption}), and end-to-end orchestration ($\S$~\ref{subsec:traces}). 

\begin{table}[t]
    \small
    \centering
    \setlength\tabcolsep{3.4pt}
    \begin{tabular}{|p{2mm}|p{27mm}|r|rr|r|r|r|} \hline
         & {\bf Application} & \multicolumn{3}{c}{{\bf Codebase}} & \multicolumn{3}{|c|}{ {\bf OCI image [MiB]}} \\ \cline{3-8} 
         & & LoC & diff & (fop) & Funky & Cont & \multicolumn{1}{|c|}{bs} \\ \hline
         \rowcolor{lightgray!40} \cellcolor{white} & \small array\_partition         &  150 &  2 & (1) & 32.7 & 1131.8 & 29.6 \\ 
                                                   & \small axi\_burst\_performance  &  126 &  8 & (7) & 66.2 & 1165.5 & 62.9 \\ 
         \rowcolor{lightgray!40} \cellcolor{white} & \small bind\_op\_storage        &   77 &  2 & (1) & 32.7 & 1131.9 & 29.7 \\ 
                                                   & \small burst\_rw                &   73 &  2 & (1) & 32.6 & 1131.7 & 29.5 \\ 
         \rowcolor{lightgray!40} \cellcolor{white} & \small critical\_path           &   92 &  4 & (3) & 36.9 & 1136.2 & 29.9 \\ 
                                                   & \small custom\_datatype         &  193 &  4 & (3) & 32.4 & 1131.7 & 29.1 \\ 
         \rowcolor{lightgray!40} \cellcolor{white} & \small dataflow\_stream         &   75 &  2 & (1) & 32.3 & 1131.4 & 29.3 \\ 
                                                   & \small dataflow\_stream\_array  &   81 &  2 & (1) & 32.4 & 1131.6 & 29.4 \\ 
         \rowcolor{lightgray!40} \cellcolor{white} & \small dependence\_inter        &   99 &  2 & (1) & 32.5 & 1131.6 & 29.4 \\ 
                                                   & \small kernel\_chain            &  195 &  6 & (5) & 33.4 & 1132.6 & 30.3 \\ 
         \rowcolor{lightgray!40} \cellcolor{white} & \small lmem\_2rw                &   76 &  2 & (1) & 32.6 & 1131.7 & 29.6 \\ 
                                                   & \small loop\_pipeline           &  119 &  2 & (1) & 32.8 & 1131.9 & 29.7 \\ 
         \rowcolor{lightgray!40} \cellcolor{white} & \small loop\_reorder            &   96 &  2 & (1) & 33.9 & 1133.0 & 30.8 \\ 
                                                   & \small partition\_cyclicblock   &  152 &  2 & (1) & 33.6 & 1132.7 & 30.5 \\ 
         \rowcolor{lightgray!40} \cellcolor{white} & \small port\_width\_widening    &  175 &  2 & (1) & 34.5 & 1133.6 & 31.4 \\ 
                                                   & \small shift\_register          &  152 &  5 & (4) & 32.9 & 1132.1 & 29.9 \\ 
         \rowcolor{lightgray!40} \cellcolor{white} & \small simple\_vadd             &  109 & 18 & (17) & 32.5 & 1131.4 & 29.5 \\ 
                                                   & \small systolic\_array          &  102 &  2 & (1) & 35.1 & 1134.2 & 32.0 \\ 
         \rowcolor{lightgray!40} \cellcolor{white} & \small wide\_mem\_rw            &   77 &  2 & (1) & 33.2 & 1132.2 & 30.0 \\ 
         \rowcolor{orange!40} \multirow{-20}{*}{\cellcolor{white}\rotatebox[origin=l]{90}{Vitis\_Accel\_Examples}} & \small common lib & 754 & 95 & (53) & - & - & - \\ \hline
        \rowcolor{lightgray!40} \cellcolor{white}  & \small 3d-rendering             & 3456 &  1 & (0) &  34.4 & 1132.3 & 29.6 \\ 
                                                   & \small digit-recognition        &  217 & 13 & (12) &  36.1 & 1134.8 & 32.5 \\ 
        \rowcolor{lightgray!40} \cellcolor{white}  & \small optical-flow             & 1624 & 75 & (74) &  60.4 & 1146.6 & 31.4 \\ 
                                                   & \small spam-filter              &  387 & 26 & (25) & 114.0 & 1213.8 & 30.7 \\ 
        \rowcolor{orange!40} \cellcolor{white} \multirow{-5}{*}{\cellcolor{white}\rotatebox[origin=l]{90}{Rosetta}} & \small common lib & 475 & 31 & (31) & - & - & - \\ \hline
        \rowcolor{lime!40} \cellcolor{white}       & \small Average                  & - & \multicolumn{2}{c|}{$3.4\%~$~($2.7\%$)} & 39.6 & 1138.2 & - \\ \hline
    \end{tabular}
    \caption{Portability study. {diff} is the total lines changed from the original code, and (fop) is only for file operations. {common lib} is code sets shared among each benchmark suite. {Funky} and {Cont} are \systemname{} unikernel and Xilinx Docker container image sizes. {bs} is the total size of bitstreams. 
    } 
    \vspace{-2mm}
    \label{tab:all_benchmark}
\end{table}

\subsection{Experimental Setup}
\label{sec:eval_setup}

\myparagraph{Testbed}
We set up four x86 servers, where three servers as worker nodes are powered by an Intel Xeon Gold 6238R Processor running at 2.2~GHz and one as a leader node powered by an Intel Xeon Gold 5317 Processor running at 3 GHz. All servers connect through a 100~Gbps switch. The worker nodes run Ubuntu 20.04, which are equipped with 256~GiB DDR4 at 2933~MHz, a 960~GiB SATA SSD as persistent storage, and an Alveo U50 FPGA via a PCIe Gen3 x16 bus. 

\myparagraph{Benchmark suites}
Since \systemname{} currently supports the OpenCL programming model, we naturally use benchmark suites written in OpenCL. We choose two benchmark suites: Vitis Accel Examples~\cite{vitis-bench} and Rosetta~\cite{rosetta}. Vitis Accel Examples is one of the most suitable benchmark suites as it is officially maintained by AMD and offers various applications to test a broad range of OpenCL APIs. We particularly use {cpp\_kernels/} of Vitis Accel Examples, containing a set of primary computation kernels, such as vector add, matrix multiplication, and an FIR filter. Rosetta is a widely used benchmark suite designed for AMD Xilinx FPGAs, offering practical, real-world FPGA workloads, including 3D rendering, digit recognition, optical flow, and spam filtering. 

\myparagraph{Baselines} 
We compare \systemname{} with two baseline setups: native execution and an industry-standard Docker container. For the container setup, we use Xilinx Base Runtime~\cite{xilinx-base-runtime}, an FPGA-accessible container officially maintained by AMD. The container runs Ubuntu 20.04 and installs a full XRT package, which directly communicates with the host-side FPGA device drivers.

\subsection{Virtualization Overheads}
\label{subsec:performance}
Firstly, we clarify \systemname{}'s FPGA virtualization overheads. 

\myparagraph{Methodology}
We measure each application's execution time in \systemname{} against the two baselines. In our experimental setup, we allocate 1~GiB of memory for each unikernel. We also break down \systemname{}'s virtualization overheads to understand their source. Specifically, we analyze the runtime overheads of OpenCL APIs, which are core FPGA operations, and initial setup overheads to create and destroy execution sandboxes, i.e., unikernels and containers. We use wide\_mem\_rw for the microbenchmarking. 

\myparagraph{Results}
Figure~\ref{fig:e2e_time} shows the end-to-end execution time of all the applications. Figures~\ref{fig:oh_funkycl} and~\ref{fig:overhead_breakdown}  show the breakdown of OpenCL API and initial setup overheads. As shown in Figure~\ref{fig:e2e_time}, \systemname{} incurs only a $7.4\%$ overhead on average compared to native execution, which is close to the container setups ($6.8\%$). We highlight that \systemname{} does not incur additional overheads for FPGA operations through OpenCL API calls as shown in Figure~\ref{fig:oh_funkycl}. The performance gaps from the baselines come from the initial setup overheads of execution sandboxes. In Figure~\ref{fig:overhead_breakdown}, for the Xilinx containers, their bootup/teardown latencies are the main factor of the performance penalties, while \systemname{} unikernels cut these overheads by 82.4\% and 83.6\%. Although we observe other setup overheads specific to \systemname{}, the main factor is not related to FPGA operations but heavy file I/O operations of IncludeOS's filesystem, causing the biggest slowdown for spam-filter (sp-fltr). Only the FunkyCL setup time, 245.1~ms on average, stems from \systemname{}'s extension, where \systemname{} copies the bitstream data for eviction/migration requests and spawns the FPGA worker thread. 

\ul{\emph{In summary}, \systemname{} induces a modest one-time setup overhead for FPGA virtualization and state management. }

\begin{figure*}[t]
    \centering
    \begin{minipage}[b]{0.33\textwidth}
      \centering
      \includegraphics[width=\linewidth]{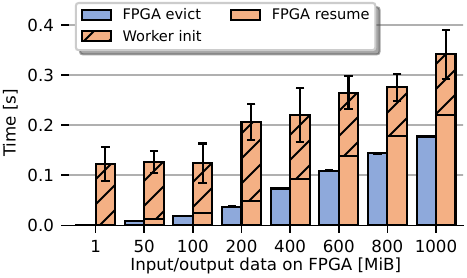}
      \caption{FPGA evict/resume overheads.}
      \label{fig:fstate_oh}
    \end{minipage} \hfill
    \begin{minipage}[b]{0.33\textwidth}
      \centering
      \includegraphics[width=\linewidth]{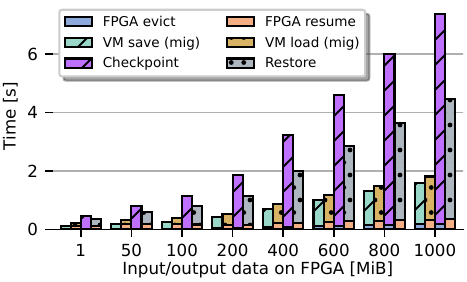}
      \caption{VM migration/checkpointing. }
      \label{fig:vmstate_oh}
    \end{minipage} \hfill
    \begin{minipage}[b]{0.33\textwidth}
      \centering
      \includegraphics[width=\linewidth]{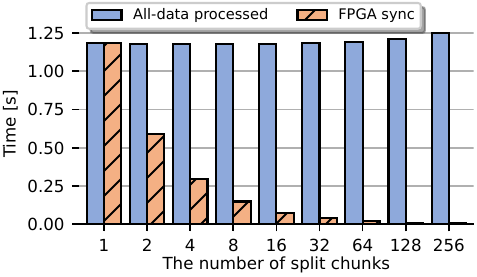}
      \caption{FPGA synchronization time.}
      \label{fig:fpga_sync_oh}
    \end{minipage}
\end{figure*}

\subsection{Application Portability}
\label{subsec:portability}
Secondly, we study \systemname{}'s application portability, i.e., how much effort developers require to port their FPGA applications. 

\myparagraph{Methodology}
We compare the Line of Code (LoC) for each application before and after porting to \systemname{}. We also analyze the OCI image sizes of \systemname{} unikernels and Xilinx Docker containers. 

\myparagraph{Results} 
Table~\ref{tab:all_benchmark} shows the summary. The total code change is just $3.4\%$ on average. We note that developers need to add just one line to the native code, {\#include <os>}, to use the OpenCL API. Most changes relate to modifying file API calls for IncludeOS, 2.7\% in total. Although IncludeOS advertises traditional file API support (e.g., fopen), we encounter various portability issues and replace file operations with the IncludeOS-specific file API (memdisk). 
Other changes are due to omitting vendor-specific OpenCL extensions~\cite{cl_ext_xilinx}.

Regarding OCI image sizes, \systemname{} unikernel is drastically smaller (28.7$\times$ on average) than the container. \systemname{}'s OCI image only contains the application binary, bitstream, input datasets, and minimal unikernel components to execute the application (e.g., FunkyCL). The bitstream and datasets are dominant in \systemname{}'s image, and the unikernel's execution binary itself is only 3-4~MiB. 
In contrast, the Xilinx Docker container significantly increases its image size due to the XRT package and its dependent libraries.

\ul{\emph{In summary}, \systemname{} enables porting OpenCL applications without changing FPGA-related code. Moreover, \systemname{}'s OCI image is substantially smaller than a vendor-provided container image, which eases to maintain their instances across distributed nodes. }


\subsection{State Management}
\label{subsec:checkpointing}
Thirdly, we evaluate \systemname{}'s state management mechanisms.

\myparagraph{Methodology}
In this experiment, we use shift\_register as a microbenchmark and measure the latency of each state management operation: FPGA eviction, VM migration, and checkpointing, while changing its input data size. Note that resuming/restoring tasks triggers FPGA reconfiguration, which takes around 3.5 seconds. We omit this overhead because it comes from a hardware limitation of the underlying Shell (Vitis XDMA), e.g., Coyote~\cite{coyote} can shorten it to 20~ms. Such hardware optimization is out of our paper's scope. 

\myparagraph{FPGA state eviction}
We first measure the time to evict and resume FPGA states. We trigger the eviction requests just after the kernel invocation is finished, ensuring that input and output (dirty) buffers exist on the FPGA memory. We change the input dataset size from 1~MB to 1000~MB. 

Figure~\ref{fig:fstate_oh} shows the results, highlighting that FPGA eviction and resumption take 177.2 and 340.8~ms for the biggest data size (1000~MiB for each input/output). 
The eviction is significantly faster for small datasets, i.e., 0.4~ms for 1~MB. 
For the eviction, \systemname{} only saves dirty buffers' states, leading to less overhead than evicting all buffers. 
{FPGA resumption} takes longer than the eviction due to (1) a consistent overhead of the worker thread creation (97.6-158.0~ms) and (2) transferring both input and output buffers. 

\ul{\emph{In summary}, \systemname{} can reasonably evict and resume FPGA states in the order of milliseconds, even for bigger datasets (1000~MiB). }

\begin{figure*}[t]
    \begin{minipage}[b]{0.33\textwidth}
        \footnotesize
        \centering
        \rowcolors{2}{}{lightgray!42}
        \setlength\tabcolsep{2.0 pt} 
        \begin{tabular}
        {|c|c|c|c|}
        \hline
            {\bf Policy} & {\bf Description} & {\bf Evict} &  {\bf Migrate} \\ \hline
            {NO\_PRE} & Reorder enqueued tasks. &  &  \\ 
            {PRE\_EV} & Evict low-priority tasks. & \cmark &  \\ 
            {PRE\_MG} & Migrate evicted tasks. &  \cmark & \cmark \\ \hline
        \end{tabular}
        \captionof{table}{Scheduling policies.}
        \label{sched_pols}
        \begin{tabular}
        {|c|c|c|c|c|}
        \hline
            {\bf App} &  {\bf Time } &  {\bf Type} & \multicolumn{2}{c|}{\bf Priority per scenario}\\ \cline{4-5}
              &  &  & Short-HP & Long-HP \\ \hline
             \rowcolor{lightgray!40} shift\_reg & 36.0 s & Long & Low & High\\
             port\_width & 35.1 s & Long & Low & High\\ 
             \rowcolor{lightgray!40} loop\_pipe & 44.8 s & Long & Low & High\\ 
             burst\_rw    & 3.5 s& Short & High & Low \\ 
             \rowcolor{lightgray!40} mem\_rw & 3.5 s & Short & High & Low\\ 
             sys\_array & 3.6 s & Short & High & Low \\  \hline
        \end{tabular}
        \captionof{table}{Execution time and priority in each scenario.}
        \label{benchs_exec}
    \end{minipage} \hfill
    \begin{minipage}[b]{0.65\textwidth}
        \begin{subfigure}[b]{0.5\linewidth}
            \centering
            \includegraphics[width=\linewidth]{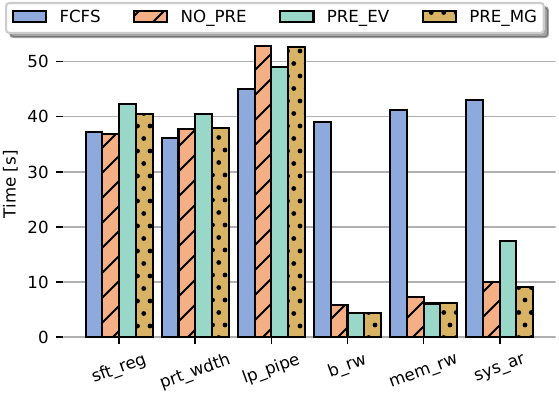}
            \vspace{-5mm}
            \caption{{\bf Short-HP} scenario.}
        \end{subfigure}
        \begin{subfigure}[b]{0.5\linewidth}  
            \centering 
            \includegraphics[width=\linewidth]{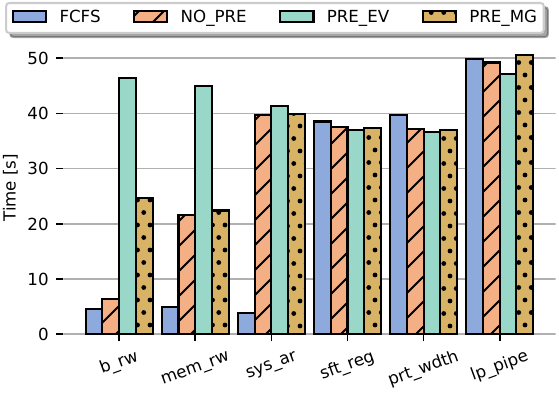}
            \vspace{-5mm}
            \caption{{\bf Long-HP} scenario.}
        \end{subfigure}
        \vspace{-5mm}
        \caption{Task preemption effectiveness. {FCFS} shows the worst case when tasks are deployed in the x-axis order, which lets low-priority tasks occupy FPGAs. }
        \label{fig:policies}
    \end{minipage}
    \vspace{-3mm}
\end{figure*}


\myparagraph{VM migration and checkpointing}
Next, we evaluate the time to save and load the entire VM state. We measure the VM saving and restoring overheads in two cases: migration (VM save, VM load), saving snapshots into host memory as a temporal cache, and checkpointing (Checkpoint, Restore), saving them into persistent storage (SSD). 

Figure~\ref{fig:vmstate_oh} shows the overhead breakdown.  
As input/output data sizes increase, the snapshot sizes also increase (125.5~MiB to 2.1~GiB). 
We observe that the overhead for each operation reasonably increases along with increasing input data size. 
Checkpoint is slower than Restore because it iterates over the whole guest memory to find dirty pages and individually writes them to the disk, leading to slower random accesses. Notably, FPGA-specific operations do not increase the overall time of VM migration and checkpointing; the FPGA eviction occupies 0.4-10.6\% in VM save and 0.1-2.4\% in Checkpoint. While the FPGA resumption gets dominant for small data sizes due to its static overheads, it gets amortized for large applications. 

\ul{\emph{In summary}, \systemname{}'s checkpoint mechanism induces reasonable overheads to make a snapshot of FPGA workloads, which enables fast recovery from system failures. }

\myparagraph{FPGA synchronization}
Lastly, we evaluate the synchronization overheads and our data splitting optimization to mitigate them  (\S~\ref{design:migration}). 
This experiment fixes the total input data at 1000~MiB and splits it into 1 to 256 chunks, measuring the overhead as the kernel is repeatedly invoked/synchronized until all the chunks are processed. 

Figure~\ref{fig:fpga_sync_oh} shows the total execution time ({All-data processed}) and FPGA synchronization time ({FPGA sync}) in the worst-case scenario, where the synchronization request is invoked immediately after the kernel invocation. 
The results highlight that our optimization can cut 96.9\% of the overhead without any performance overheads (<0.1\%) in the case of 32 chunks, where each kernel invocation processes 31.25~MiB data. We also observe that  excessive data splitting increases the total time, e.g., 5.5\% with 256~chunks. \systemname{} can prevent it by configuring the lower boundary of the chunk size. 

\ul{\emph{In summary}, the results demonstrate that our optimization can effectively eliminate the waiting time for FPGA synchronization. }

\begin{figure*}[t]
    \centering
    \begin{minipage}[]{0.33\textwidth}
        \centering
        \includegraphics[width=\linewidth]{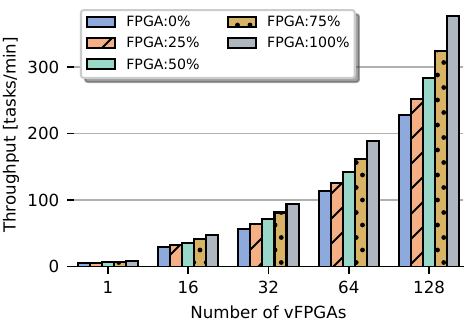}
        \caption{Workload scaling. }
        \label{fig:scalev}
    \end{minipage} \hfill
    \begin{minipage}[]{0.33\textwidth}
      \centering
      \includegraphics[width=\linewidth]{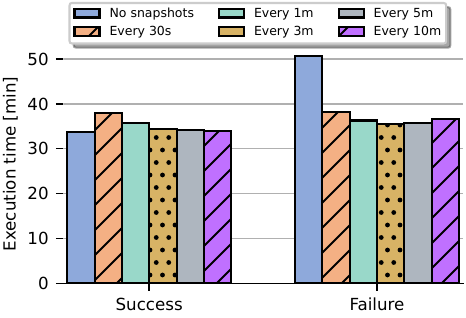}
      \caption{Fault tolerance analysis. }
      \label{fig:fault_tolerance}
    \end{minipage} \hfill
    \begin{minipage}[]{0.33\textwidth}
      \centering
      \includegraphics[width=\linewidth]{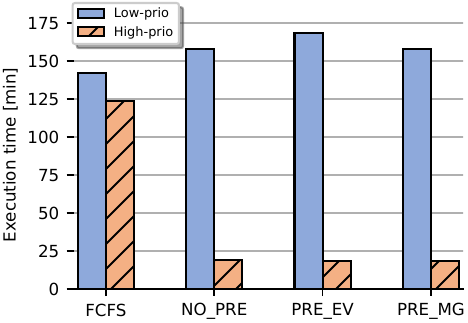}
      \caption{Preemptive task scheduling.}
      \label{fig:sched_sim}
    \end{minipage} \hfill
    \vspace{-4mm}
\end{figure*}



\subsection{Task Preemption}
\label{subsec:preemption}
Next, we examine how \systemname{}'s state management mechanisms are effective for preemptive task scheduling on our FPGA cluster. 

\myparagraph{Methodology}
We execute our orchestrator's prototype on the leader node and the \systemname{} runtime daemons in the other three worker nodes.
Table~\ref{sched_pols} shows the priority-based scheduling policies we compare. Every policy deploys tasks in the scheduling queue in a First-Come-First-Served (FCFS) manner as long as any FPGA is free. If all the FPGAs are occupied by running tasks, {NO\_PRE} sorts the waiting queue based on the task's priority. On the other hand, {PRE\_EV} evicts a low-priority task when a higher-priority task comes. {PRE\_MG} performs both the eviction and migration; the migration takes place when an evicted task can be resumed on a different node.

We create two scenarios where we assign priorities to each task considering its execution time: {\bf 1. Short-HP} prioritizes short-lived tasks labeled as high priority (HP), and {\bf 2. Long-HP} prioritizes long-running tasks. We use a subset of Vitis Accel Examples, which contains six applications with long and short execution times, as shown in Table~\ref{benchs_exec}. We perform the batch execution in both scenarios, repeatedly sending deployment requests of the six applications while changing their orders to cover all possible deployment orders of high- and low-priority tasks. We test 20 patterns of different orders and measure the average execution time for each benchmark. 

\myparagraph{Results}
Figure~\ref{fig:policies}~(a) shows the behavior of the different policies in the Short-HP scenario, where {PRE\_MG} is most effective for high-priority tasks, reducing 16.7\% of their average execution time compared to {NO\_PRE}. In the case of non-preemptive policies ({FCFS} and {NO\_PRE}), we observe that high-priority tasks ({b\_rw}, {mem\_rw}, {sys\_ar}) take longer execution time because they have to wait for the completion of low-priority tasks when all three FPGAs are in use. On the other hand, when preemption is enabled ({PRE\_EV} and {PRE\_MG}), the long-running tasks get evicted in favor of high-priority tasks. There is one edge case, sys\_ar for PRE\_EV, whose deployment is impeded by long FPGA synchronization of lp\_pipe consisting of only two heavy FPGA operations. Such a case can be mitigated by our optimization demonstrated in Figure~\ref{fig:fpga_sync_oh}. 

Figure~\ref{fig:policies}~(b) shows the Long-HP scenario, where {PRE\_EV} is most effective for high-priority tasks: 2.2\% faster than {NO\_PRE}. {PRE\_MG} is less effective because evicted low-priority tasks are sometimes migrated before high-priority tasks arrive, slightly delaying their deployment. This is a common issue of task preemption, which can be mitigated by adapting better algorithms~\cite{kairos,neptune,prema}. 

\ul{\emph{In summary}, the evaluation demonstrates that \systemname{} enables FPGA task preemption across distributed FPGA nodes, leading to shorter execution times than non-preemptive scheduling. }

\subsection{End-to-end Orchestration}
\label{subsec:traces}
Lastly, we evaluate the effectiveness of \systemname{}'s orchestration services. Due to our limited number of FPGA resources, we simulate a large-scale FPGA cluster using production traces from Google. 

\myparagraph{Methodology} 
We use production traces from the Google ClusterData 2019 dataset~\cite{borg-trace-v2019}. The dataset comprises one month of real-world application traces (execution, failure, and eviction) from eight Google Borg clusters~\cite{borg,borg-2nd}. 

Because the traces only refer to CPU workloads, we strive for trace modification to reasonably simulate the behavior of FPGA-accelerated jobs. First, we manipulate the execution time of each trace job based on the expected performance gains of FPGA acceleration against CPUs. Specifically, we calculate the performance gains by comparing the average execution time of Rosetta real-world applications on the U50 FPGA and CPU, as Rosetta offers both CPU-only and FPGA implementations. As a result, we observe that FPGAs are 1.6$\times$ faster than CPUs on average. We apply this speedup factor to the job's execution time. Second, to simulate state management operations, we estimate the FPGA memory usage of each job based on its CPU memory usage. While CPU jobs can offload any portion of their computation to FPGAs in theory, we assume the worst-case scenario for \systemname{} for fairness, where all computations are offloaded to FPGAs, and all FPGA memory buffers need to be saved/restored, leading to the biggest checkpointing overheads. As input/output buffers are duplicated on both CPU and FPGA memory, we assume that FPGA memory usage is always the same as the CPU but limited to its physical memory capacity, i.e., 8~GiB on U50. 


Using the modified traces, we implement a full-system orchestration simulator written in Python. The simulator maintains the state of an FPGA-equipped cluster that offers a fixed number of resources (CPUs, memory, vFPGAs) and replays the execution of jobs as they appear in the traces. During the simulation, the simulator parses the traces to retrieve timestamps of scheduling events that happened to tasks: task submission, execution, eviction, failure, and completion. Whenever any events happen, the simulator updates the cluster state and accordingly inserts the \systemname{}-specific overheads, e.g., unikernel bootups.

\myparagraph{Scalability}
We first evaluate how \systemname{}'s performance scales with the number of vFPGAs in the cluster. Assuming that all types of workloads cannot be fully accelerated by FPGAs, we define \textit{acceleration rates}, which indicate the duration tasks can be accelerated with FPGAs during their execution periods. We increase the cluster size from $1$ to $128$ vFPGAs for five different acceleration rates: 0\% (no FPGA acceleration), 25\%, 50\%, 75\%, and 100\% (fully accelerated). For each cluster size, we simulate the execution of all tasks in the traces and calculate the throughput, i.e., completed tasks per minute. 

Figure~\ref{fig:scalev} shows the results. We observe that an increasing number of vFPGAs and acceleration rates reasonably improve the system throughput. We note that even a small acceleration rate (25\%) achieves 1.1$\times$ higher throughputs than no FPGA acceleration (0\%), indicating that \systemname{}'s virtualization overheads are sufficiently small to retain FPGA's performance benefits for real-world traces. 

\ul{\emph{In summary}, \systemname{} scales effectively with cluster size on real-world application traces, despite virtualization overheads.} 

\newcommand{\xred}{\textcolor{red}{\xmark}}
\newcommand{\cgreen}{\textcolor{teal}{\cmark}}
\begin{table*}[t]
    \small
    \centering
    \begin{tabular}{|c|c|c|c|c|c|c|c|c|c|} \hline
        {\bf Solutions} & \multicolumn{2}{c|}{{\bf FPGA virtualization and isolation}} & \multicolumn{3}{c|}{{\bf FPGA state management}} & \multicolumn{4}{c|}{{\bf Cloud-native orchestration}} \\ \cline{2-10}
                                 &  {\bf Guest sandbox} & {\bf FPGA fabric} & {\bf Guest} & {\bf Logic} & {\bf Memory} & {\bf OCI/CRI} & {\bf Scheduling} & {\bf Checkpoint} & {\bf Scaling} \\ \hline
        \rowcolor{lightgray!40} \multicolumn{10}{|c|}{\textit{Industry practices}} \\
        Cloud instance~\cite{ec2f2}                 & VM (standard)  & N/A                   & -       & -                  & -                  & -             & -       & -       & -       \\ 
        k8s plugin~\cite{xilinx-k8s,intel-fpga-k8s} & Container      & Vendor's Shell$^\ast$ & -       & -                  & -                  & \cgreen       & \cgreen & -       & -       \\ \hline
        \rowcolor{lightgray!40} \multicolumn{10}{|c|}{\textit{State-of-the-art research}} \\
        AmorphOS~\cite{amorphos} & User process   & Own Shell             & -       & -                  & -                  & -             & -       & -       & -       \\ 
        Coyote~\cite{coyote} & User process   & Own Shell             & -       & -                  & -                  & -             & -       & -       & -       \\ 
        SYNERGY~\cite{synergy}                      & User process   & AmorphOS$^\ast$       & -       & \cgreen            & -                  & -             & -       & -       & -       \\ 
        Optimus~\cite{optimus}                      & VM (standard)  & Own Shell             & -       & -                  & -                  & -             & -       & -       & -       \\ 
        AvA~\cite{ava}                              & VM (standard)  & AmorphOS$^\ast$       & \cgreen & \cgreen$^\diamond$ & \cgreen$^\diamond$ & -             & -       & -       & -       \\ 
        BlastFunction~\cite{blastfunction}          & Container      & Vendor's Shell$^\ast$ & -       & -                  & -                  & \cgreen       & -       & -       & \cgreen \\ 
        \rowcolor{lime!30}
        {\bf \textcolor{teal}{Funky}} & {\bf \textcolor{teal}{VM (unikernel)}} & Vendor's Shell$^\ast$ & \cgreen & (\cgreen) & \cgreen & \cgreen & \cgreen & \cgreen & \cgreen \\ \hline
    \end{tabular}
    \begin{tablenotes}
        \item[*]\hspace{3mm}$\ast$: leverages existing FPGA OSes (Shells) to isolate FPGA fabrics. \hspace{13mm} $\diamond$: adopts \textit{record-and-replay}~\cite{10.1145/2492705}, which does not directly save the states. 
    \end{tablenotes}
    \caption{Comparison with the state-of-the-art approaches for leveraging FPGAs in cloud environments. }
    \vspace{-4mm}
    \label{tab:fpga_virt_studies}
\end{table*}

\myparagraph{Fault tolerance}
We next evaluate the effectiveness of \systemname{}'s checkpointing mechanism for fault-tolerant execution. During the simulation, the system periodically takes snapshots of running tasks, while all the tasks fail at random points during their execution (1-99\%). Under this setup, the failure events happen at 50\% of the total execution time on average, which roughly follows the real-world scenario reported by a prior study~\cite{7980073}, i.e., failed cluster jobs run for roughly 40\% of their total execution time before the first task failure event. Whenever a task fails, the system resumes the task from the latest snapshot or restarts from the beginning if there are no snapshots. We also examine checkpointing overheads by simulating the case where no failure happens (Success). 

Figure~\ref{fig:fault_tolerance} shows the average execution times for different checkpointing durations. In any duration, checkpointing reduces the execution time of failed tasks by recovering from the latest snapshot. Regarding the checkpointing duration, frequent checkpointing (e.g., every 30 seconds) increases the performance overheads, while sparse checkpointing (e.g., every 10 minutes) mitigates the overheads in success at the cost of longer recovery time in failure. 

\ul{\emph{In summary}, \systemname{} 
recovers failed task performance using snapshots when the snapshot frequency is optimized.}

\myparagraph{Scheduling}
Lastly, we examine the effectiveness of our preemptive task scheduling for the trace jobs. We evaluate the same scheduling policies used in $\S$~\ref{subsec:preemption} with 32 vFPGAs. 

Figure~\ref{fig:sched_sim} shows the results, which represent the same trend as Figure~\ref{fig:policies}. \systemname{}'s preemptive policies, PRE\_EV and PRE\_MG, reduce the execution time of high-priority tasks compared to NO\_PRE, 5.3\% and 4.5\% shorter. PRE\_MG also reduces 5.9\% of the execution time of low-priority tasks compared to PRE\_EV because the FPGA is more likely to be occupied for long-running jobs, and task eviction and migration become more effective. 

\ul{\emph{In summary}, \systemname{}'s preemption mechanism is effective for the production traces as well as the real hardware evaluation. } 

\section{Related Work}  
\label{related-research}
We compare \systemname{} with state-of-the-art research on cloud FPGA management across three dimensions, summarized in Table~\ref{tab:fpga_virt_studies}. 

\myparagraph{FPGA virtualization and isolation}
Vendor-provided FPGA platforms adopt the PCIe passthrough that statically binds FPGA devices to guest VMs~\cite{when-fpga-meets-cloud-tcc2020,ec2f2,azure-np}. While these dedicated instances achieve near-zero FPGA control overheads, the lack of FPGA virtualization leads to low resource utilization in a multi-tenant cloud. FPGA virtualization based on {\em FPGA OS} applies the OS primitives to hardware tasks on FPGAs~\cite{ecos,hthreads,reconos,borph,virt_hwos}. Task schedulers~\cite{6176537,4724892,6861637,1336761,6861366}, memory virtualization~\cite{leap_scratchpads,coram,matchup}, security~\cite{tdos-apsys2023,cloud-fpga-security-2020}, communication layers~\cite{virtiofpga,5170181,vfpio,semperos,rack-scale-ukernel-apsys2020} have been actively studied. AmorphOS~\cite{amorphos} and Coyote~\cite{coyote} aim to isolate and share FPGA fabric on a single board rather than cloud-scale distributed FPGA orchestration. \systemname{} 
leverages such FPGA OSes or vendor-provided Shells~\cite{vitis-platform,intel-afu} to isolate the FPGA fabric and onboard devices (e.g., memory) while it strives to offer a hypervisor-level isolated sandbox for guest CPU applications, which is crucial in a multi-tenant cloud. 

Optimus~\cite{optimus} and AvA~\cite{ava} offer hypervisor-level isolation for standard VMs on traditional cloud FPGA platforms (e.g., Amazon F2~\cite{ec2f2}). Although they achieve \systemname{}'s security level, they do not consider integration with cloud-native orchestration. In addition, large contexts of standard VMs do not fit cloud-native applications, leading to slow boot time and performance penalties. While AvA provides a VM migration mechanism, integrating it with orchestrators is not its primary goal. In contrast, \systemname{} is the first work to leverage lightweight sandboxes (i.e., unikernels) for FPGA virtualization and integrate the FPGA state management operations into an industry-standard orchestrator, i.e., Kubernetes.

\myparagraph{FPGA state management}
FPGA context switching~\cite{Lee2010HardwareCM,5290950,1515726,903426} and checkpointing~\cite{Knodel,HDLmig,10.5555/647927.739382,7284295,reconos-multi} have been mainly exploited at the hardware level.  They are either based on reading back the state of FPGA logic (registers, BRAMs) or modifying the hardware embedding scan chains to extract the state~\cite{reconos-multi}. SYNERGY~\cite{synergy} is a compiler-based approach that transforms the HDL code of user logic into a preemptible design. While these approaches focus on user logic states, \systemname{} comprehensively breaks down the FPGA workload states in the state-of-the-art architecture (logic, memory, VM) and introduces an end-to-end flow to save and restore them. AvA~\cite{ava} adopts \textit{record-and-replay}~\cite{10.1145/2492705}, which can induce significant overheads for long-running services.

\myparagraph{Cloud-native orchestration} 
Many orchestration engines have been developed and studied for efficient resource management in data centers and commercial clouds~\cite{kubernetes,borg,10433234}, offering preemptive scheduling~\cite{omega,10.1145/3369583.3392671,10214311}, checkpointing~\cite{10.1145/3361525.3361535,electronics10040423}, fault tolerance~\cite{HASAN2018156}, workload scaling~\cite{borg-2nd,s20164621,9732997}. However, they target only CPU and memory resources. GPU orchestration is being rapidly exploited~\cite{kube-knots,streambox,288717} due to the emergence of AI workloads. \systemname{} tackles the orchestration of FPGAs, given their architectural difference from CPUs and GPUs. 

There are a few projects to leverage FPGAs in cloud-native environments, which partially address our goals. BlastFunction~\cite{blastfunction} and F3~\cite{f3} integrate FPGAs into Kubernetes-based serverless frameworks, while their functionality is restricted to workload scaling and non-preemptive scheduling. 
Molecure~\cite{molecule} also supports FPGA acceleration but lacks orchestration and state management support. Unlike these serverless studies, \systemname{} offers comprehensive FPGA virtualization and state management, enabling broader orchestration services like migration, preemption, and checkpointing. 

\section{Conclusion}
We present \systemname{}, an FPGA orchestration engine for cloud-native environments with the following contributions. 
First, we design {\em a unikernel architecture for FPGA virtualization} that not only minimizes the performance penalties but also brings faster boot times than containers.  We also design a guest OpenCL-compatible library for application portability. Second, we present {\em a hypervisor-driven FPGA state management}. The thin hypervisor transparently traces host-FPGA data transfers and enables saving and restoring unikernel/FPGA contexts.  Lastly, we present three orchestration services: preemptive scheduling, checkpointing, and workload scaling. Our orchestration mechanism is {\em compatible with the industry-standard CRI/OCI specifications}. 
We implement and evaluate \systemname{} on our four-node FPGA cluster with three AMD Xilinx FPGAs.
Our evaluation demonstrates that \systemname{} imposes virtualization overheads of $7.4\%$ against native execution while enabling FPGA virtualization and state management for orchestration operations.  
\myparagraph{Artifact availability}
The \systemname{} codebase is publicly available at \url{https://github.com/TUM-DSE/Funky.git}. 

\myparagraph{Supplements}
Our appendix includes a discussion about \systemname{}'s potential applicability for other applications and system domains. 

\subsection*{Acknowledgements}
This work was supported in parts by an ERC Starting Grant (ID: 101077577) and the Chips Joint Undertaking (JU), European Union (EU) HORIZON-JU-IA, under grant agreement No. 101140087 (SMARTY).
The authors acknowledge the financial support by the Federal Ministry of Research, Technology and Space of Germany in the programme of “Souverän. Digital. Vernetzt.”. Joint project 6G-life, project identification number: 16KISK002

\appendix
\section{Appendix}

\subsection{Discussion}
\label{sec:discussion}

We discuss \systemname{}'s potential applicability for other applications and system domains. 

\myparagraph{Applicability to other FPGA workloads}
Our evaluation includes four real-world FPGA workloads from Rosetta~\cite{rosetta}, categorized as graph processing (digit recognition), image processing (3d rendering and optical flow), and pattern matching (spam filtering). Besides Rosetta, \systemname{} is capable of adapting other recent, practical FPGA workloads such as machine learning inference (PipeCNN~\cite{pipecnn}, HLS4ML~\cite{hls4ml}), database query processing (SQL2FPGA~\cite{sql2fpga}), data compression~\cite{vitis-comp-lib}, and graph processing~\cite{vitis-graph-lib} because they are also designed as coprocessor-like kernels for streaming computation. Their execution and control flow are the same as the host-driven FPGA execution model primarily supported by \systemname{}. 

\myparagraph{FPGA space sharing}
Our \systemname{}'s prototype does not currently support space sharing (i.e., a single FPGA has multiple vFPGAs) because the Xilinx Runtime (XRT) provides only one reconfigurable slot. Therefore, for FPGA memory management and isolation, \systemname{} simply ensures that only one application is running on each FPGA, and once the task is complete, the corresponding \systemname{} monitor flushes FPGA memory contents before a new application acquires it. \systemname{} can be extended for a space-sharing scenario by adapting multi-slot FPGA platforms such as Coyote~\cite{coyote,coyote-v2}, which offers hardware MMUs dedicated to each vFPGA and ensures memory isolation between vFPGAs.

\myparagraph{Saving and restoring FPGA logic states}
Besides coprocessor-like kernels, FPGAs are used to deploy non-streaming compute units (e.g., soft CPU cores~\cite{riscv}) that keep their contexts in the on-chip memory (BRAM) and registers, whose states are inaccessible from the host. Although such stateful kernels are not the primary targets of our state management mechanism, \systemname{} can be extended to support them by leveraging existing FPGA checkpointing techniques such as StateReveal~\cite{statereveal} and SYNERGY~\cite{synergy}. These techniques perform a static code analysis for the kernel’s HDL code, detecting state elements (registers, BRAMs) and automatically integrating state-capture circuits into the kernel, which allows \systemname{} to safely pause/restart the kernel execution, and save/restore all internal states via control interfaces (e.g., AXI control buses). We need a small hardware modification to expose their control interfaces to the \systemname{} monitor as memory-mapped registers, which can be realized by a simple HDL wrapper module for each application's kernel. 

\myparagraph{Quality of Services (QoS)}
Our \systemname{} orchestrator primarily aims to demonstrate that \systemname{}’s eviction/migration features can be integrated into industry-standard orchestrators, rather than demonstrating end-to-end QoS guarantees. However, our evaluation includes the effectiveness of \systemname{}’s preemptive task scheduling, partially demonstrating the fairness of \systemname{}. The evaluation also demonstrates that \systemname{} can be as fair as common priority-based schedulers, as it allows multiple tasks with the same priority to evenly share FPGA resources by task preemption.

\bibliographystyle{plain}
\bibliography{references}


\end{document}
\endinput